# Formulation of a Family of Sure-Success Quantum Search Algorithms

Jin-Yuan Hsieh[1], Che-Ming Li[2], Jenn-Sen Lin[3], and Der-San Chuu[3]

[1]Department of Mechanical Engineering, Ming-Hsin University of Science and Technology, Hsinchu 30441, Taiwan

[2] Institute and Department of Electrophysics, National Chiao Tung University, Hsinchu 30050, Taiwan

[3] Department of Mechanical Engineering, Lien-Ho Institute of Technology, MaioLi 36012, Taiwan

## Abstract

In this work, we consider a family of sure-success quantum algorithms, which is grouped into even and odd members for solving a generalized Grover search problem. We prove the matching conditions for both groups and give the corresponding formulae for evaluating the iterations or oracle calls required in the search computation. We also present how to adjust the phase angles in the generalized Grover operator to ensure the sure-success if minimal oracle calls are demanded in the search.

Since developed, the Grover quantum search algorithm [1] has received many attentions because it provides a quadratic speedup over its counterpart. If there is one item among $N$ unsorted items to be searched, then using the Grover algorithm will accomplish the computation in $O(\sqrt{N})$ quantum mechanical steps, instead of $O(N)$ classical steps. The Grover algorithm is carried out by successively applying the Grover operator on an initial state, which is usually prepared as a uniform superposition of all states, to amplify the probability amplitude of the marked state. The Grover operator is composed of the unitary transformations of the $\pi$-inversion of the marked state and the $\pi$-inversion about average. Multi-object problems can also be solved by the Grover algorithm, and in this case the marked state will be the uniform superposition of all target states. Although the Grover algorithm has been proved to be optimal [2] in the sense that it requires minimal oracle calls to accomplish the computation, it in face provides a high probability of finding the marked state only for a large $N$. The probability will be lower as $N$ decreases. Hsieh and Li [3] and Long [4], however, have shown a sure-success quantum search algorithm acceptable for arbitrary $N$, provided that the transformations of the $\pi$-inversion of the marked state and the $\pi$-inversion about average are replaced by generalized $\phi-$ and $\theta-$ relations, respectively, and that the matching condition

$\phi = \theta$ is obeyed. That is, if we denote $I_\tau = 1 + (e^{i\phi} - 1)|\tau\rangle\langle\tau|$ and $I_s = -1 + (1 - e^{i\theta})|s\rangle\langle s|$, where $|\tau\rangle$ and $|s\rangle$ are the marked and initial state, respectively, then the generalized Grover operator is given by $G = I_s I_\tau$ and in $n$ iterations the unit probability for finding the marked state is ensured if $\phi = \theta$, viz., we will have $p = |\langle\tau|G^n|s\rangle|^2 = 1$.

Instead of applying $G^n$ on initial state, Hu [5] recently introduced other choices of sure-success algorithm. He presented the use of the operators $A_{2n} = (I_s^\dagger I_\tau^\dagger I_s I_\tau)^n$ and $A_{2n+1} = GA_{2n}$ to accomplish sure-success search computations and named the former the even number and the latter the odd number of the family $\{A_n, n=1,2,\cdots\}$ because they require even ($2n$) and odd ($2n+1$) oracle calls, respectively. In his paper [5], Hu predicted, without proofs, the matching condition $\phi = -\theta$ for the algorithm applying the even member $A_{2n}$ and the condition $\phi = \theta$ for that using the odd member $A_{2n+1}$, but has only discussed the cases using the members $A_1$, $A_2$, $A_4$ and $A_6$, and derived the corresponding relations between the phase angle $\theta$ and the fraction of all targets in the unsorted data base, namely, $M/N$, where $M$ is the number of the targets. (In the paper [5], Hu actually used the phase angles $\phi + \pi$ and $(\pi - \theta)/2$. Clearly, it is impossible to analyze every member of the family $\{A_n, n=1,2,\cdots\}$, following the same way that Hu has implemented. In this work, however, we will show that the sure-success algorithms considered in ref. [5], like the generalized Grover algorithm derived in refs. [3] and [4], can be fully formulated. We will prove the matching conditions that Hu predicted for both the algorithms using the even and odd members, respectively, and give the detail formulae for evaluating the required oracle calls as functions of the phase angle $\theta$ (or $\phi$) and the given initial probability amplitude of the marked state.

Consider a two-dimensional, complex Hilbert space spanned by the marked state $|\tau\rangle$ and the state $|\tau_\perp\rangle$, which is orthogonal to $|\tau\rangle$. The initial state, as a uniform superposition of all states, then can be expressed by

$$|s\rangle = \sin(\beta)|\tau\rangle + \cos(\beta)|\tau_\perp\rangle, \qquad (1)$$

where $\sin\beta \equiv \sqrt{M/N}$. The two sure-success quantum search algorithms using the even member $A_{2n}$ and the odd member $A_{2n+1}$ are considered. In both cases we will require $\langle\tau_\perp|A_{2n}|s\rangle = 0$ and $\langle\tau_\perp|A_{2n+1}|s\rangle = 0$, respectively, to ensure the sure-success in finding the marked state $|\tau\rangle$.

In matrix form, the operator $I_s^\dagger I_\tau^\dagger I_s I_\tau$ can be written

$$I_s^\dagger I_\tau^\dagger I_s I_\tau = \begin{bmatrix} \cos(w) + i\sin^2(\tfrac{\theta}{2})\sin(\phi)\sin^2(2\beta) & 2r\sin(\tfrac{\theta}{2})\sin(\tfrac{\phi}{2})\sin(2\beta)e^{-i(\tfrac{\phi}{2}-\gamma)} \\ -2r\sin(\tfrac{\theta}{2})\sin(\tfrac{\phi}{2})\sin(2\beta)e^{i(\tfrac{\phi}{2}-\gamma)} & \cos(w) - i\sin^2(\tfrac{\theta}{2})\sin(\phi)\sin^2(2\beta) \end{bmatrix}, \quad (2)$$

where the notations $w$, $r$ and $\gamma$ are defined by

$$\cos(w) = 1 - 2\sin^2(\tfrac{\theta}{2})\sin^2(\tfrac{\phi}{2})\sin^2(2\beta) \qquad (3)$$

and

$$re^{i\gamma} = \cos(\tfrac{\theta}{2}) + i\sin(\tfrac{\theta}{2})\cos(2\beta). \qquad (4)$$

Note that since the above matrix has a determinant of unity, we thus have the useful relation

$$\sin^2(w) = [\sin^2(\tfrac{\theta}{2})\sin(\phi)\sin^2(2\beta)]^2 + [2r\sin(\tfrac{\theta}{2})\sin(\tfrac{\phi}{2})\sin(2\beta)]^2. \qquad (5)$$

The eigenvalues of the operator $I_s^\dagger I_\tau^\dagger I_s I_\tau$ are $\lambda_{1,2} = e^{\pm iw}$ and the corresponding eigenstates are computed

$$|\lambda_1\rangle = \cos(x)|\tau\rangle + i\sin(x)e^{i(\tfrac{\phi}{2}-\gamma)}|\tau_\perp\rangle \text{ and } |\lambda_2\rangle = i\sin(x)e^{-i(\tfrac{\phi}{2}-\gamma)}|\tau\rangle + \cos(x)|\tau_\perp\rangle. \qquad (6)$$

In expression (6), the notation $x$ is defined by

$$\sin(x) = 2r\sin(\tfrac{\theta}{2})\sin(\tfrac{\phi}{2})\sin(2\beta)/\sqrt{\ell} \text{ or}$$

$$\cos(x) = [\sin(w) + \sin^2(\tfrac{\theta}{2})\sin(\phi)\sin^2(2\beta)]/\sqrt{\ell}, \qquad (7)$$

where

$$\ell = [\sin(w) + \sin^2(\tfrac{\theta}{2})\sin(\phi)\sin^2(2\beta))]^2 + (2r\sin(\tfrac{\theta}{2})\sin(\tfrac{\phi}{2})\sin(2\beta))]^2$$

$$= 2\sin(w)[\sin(w) + \sin^2(\tfrac{\theta}{2})\sin(\phi)\sin^2(2\beta)].$$

Once the eigenvalues $\lambda_{1,2}$ and the corresponding eigenstates $|\lambda_{1,2}\rangle$ are determined, then the operator $A_{2n}$ can be simply expressed by the spectral decomposition $A_{2n} = e^{inw}|\lambda_1\rangle\langle\lambda_1| + e^{-inw}|\lambda_2\rangle\langle\lambda_2|$. In matrix form, the operator $A_{2n}$ can also be written

$$A_{2n} = \begin{bmatrix} \cos(nw) + i\sin(nw)\cos(2x) & \sin(2x)\sin(nw)e^{-i(\frac{\phi}{2}-\gamma)} \\ -\sin(2x)\sin(nw)e^{i(\frac{\phi}{2}-\gamma)} & \cos(nw) - i\sin(nw)\cos(2x) \end{bmatrix}. \qquad (8)$$

Now, it is easy to show that in the case when the even member $A_{2n}$ is used, the requirement $\langle \tau_\perp | A_{2n} | s \rangle = 0$ leads to

$$\sin(2x)\sin(\beta)\sin(\tfrac{\phi}{2}-\gamma) + \cos(2x)\cos(\beta) = 0, \qquad (9)$$

and

$$\cos(nw)\cos(\beta) - \sin(nw)\sin(2x)\sin(\beta)\cos(\tfrac{\phi}{2}-\gamma) = 0. \qquad (10)$$

By the definition of x, we deduce the relation
$$\frac{\sin(2x)}{\cos(2x)} = \frac{r}{\sin(\tfrac{\theta}{2})\cos(\tfrac{\phi}{2})\sin(2\beta)} \qquad (11)$$

Then, as the relation (11), incorporated with the definition (4), is used equation (9) will reduce to

$$\frac{\cos(2x)\sin(\beta)}{\sin(\tfrac{\theta}{2})\cos(\tfrac{\phi}{2})\sin(2\beta)} \sin(\tfrac{\phi+\theta}{2}) = 0. \qquad (12)$$

So we have the first matching condition $\phi = -\theta$, which is exactly the one predicted by Hu [5], for the sure-success algorithm using the even member $A_{2n}$. By the relation in (11) and the definition of $r$, we can further compute

$$\sin(2x) = \left( \frac{1 - \sin^2(\tfrac{\theta}{2})\sin^2(2\beta)}{1 - \sin^2(\tfrac{\theta}{2})\sin^2(\tfrac{\phi}{2})\sin^2(2\beta)} \right)^{1/2}.$$

Consequently, expression (10) then reduces to
$$\cos(nw - \delta_e) = 0, \qquad (13)$$

where $\delta_e = \sin^{-1}[\sin(\beta)\cos(\phi/2-\gamma)]$. Using the matching condition $\phi = -\theta$, we thus derive the exact formula for evaluating the required iterations in the sure-success algorithm using $A_{2n}$,

$$n_e(\beta,\theta) = \lceil f_e(\beta,\theta) \rceil, \qquad (14)$$

where the symbol $\lceil\ \rceil$ denotes the smallest integer greater than the quantity in it, and $f_e$ is given by

$$f_e(\beta,\theta) = \frac{\tfrac{\pi}{2} + \sin^{-1}\left( \frac{\sin(\beta)[1 - 2\sin^2(\tfrac{\theta}{2})\cos^2(\beta)]}{\sqrt{1-\sin^2(\tfrac{\theta}{2})\sin^2(2\beta)}} \right)}{\cos^{-1}[1 - 2\sin^4(\tfrac{\theta}{2})\sin^2(2\beta)]}.$$

Similarly, in the sure-success algorithm using the odd member $A_{2n+1}$, the

requirement $\langle\tau_\perp|A_{2n+1}|s\rangle = 0$ will give

$$\cos(nw)\sin(\tfrac{\theta-\phi}{2})\cos(\beta) + \\ \sin(nw)[-\cos(2x)\cos(\tfrac{\theta-\phi}{2})\cos(\beta) + \sin(2x)\cos(\gamma)\sin(\tfrac{\theta}{2})\sin(2\beta)\cos(\beta)] = 0 \quad (15)$$

and

$$\cos(nw)[\cos(\tfrac{\theta-\phi}{2}) - 4\sin(\tfrac{\theta}{2})\sin(\tfrac{\phi}{2})\sin^2(\beta)]\cos(\beta) \\ + \sin(nw)[\cos(2x)(\sin(\tfrac{\theta-\phi}{2}) - 4\sin(\tfrac{\theta}{2})\cos(\tfrac{\phi}{2})\sin^2(\beta)]\cos(\beta) \quad (16) \\ - \sin(2x)[\cos(\gamma + \tfrac{\theta}{2}) + 4\sin(\gamma)\sin(\tfrac{\theta}{2})\cos^2(\beta)]\sin(\beta) = 0.$$

Again, as the relation in (11) and the definition of $r$ are used, equation (15) will reduce to

$$\cos(\beta)\sin(\tfrac{\theta-\phi}{2})\left(\cos(nw) - \sin(nw)\frac{\cos(2x)\sin(\tfrac{\phi}{2})}{\cos(\tfrac{\phi}{2})}\right) = 0 \quad (17)$$

One thus obtains the second matching condition $\phi = \theta$ for the algorithm using the odd member $A_{2n+1}$, consistent with what Hu [5] expected.

Substituting the condition $\phi = \theta$ into (16), one eventually deduces the formula for evaluating the required iterations in the sure-success algorithm using the odd member $A_{2n+1}$,

$$n_o(\beta,\theta) = \lceil f_o(\beta,\theta) \rceil \quad (18)$$

where $f_o$ is given by

$$f_o(\beta,\theta) = \frac{\tfrac{\pi}{2} - \cos^{-1}\left(\dfrac{\cos(\beta)[1-4\sin^2(\tfrac{\theta}{2})\sin^2(\beta)]\sqrt{1-\sin^4(\tfrac{\theta}{2})\sin^2(2\beta)}}{\sqrt{1-\sin^2(\tfrac{\theta}{2})\sin^2(2\beta)}}\right)}{\cos^{-1}[1-2\sin^4(\tfrac{\theta}{2})\sin^2(2\beta)]}$$

Note that in this case the inequality $1-\sin^4(\theta/2)\sin^2(2\beta) \geq 0$ should be demanded since then the meaningful requirement $f_o \geq 0$ can thus be fulfilled. Expressions (14) and (18) indicate that as $n_e$ or $n_o$ is designated arbitrarily, the corresponding relation between θ and β can be determined by these formulae. In ref.[5], Hu has only discussed the cases of $n_o = 0$ and $n_e = 1, 2$, and 3, while in this work, we actually have fully formulated the whole family { $A_n$, $n=1,2,\cdots$ }, and proved the matching conditions that Hu predicted. In below, we give a brief discussion on the present formulae.

Let us denote the oracle-call functions by $c_e(\beta,\theta) = 2f_e(\beta,\theta)$ for the algorithm using the even member $A_{2n}$ and $c_o(\beta,\theta) = 2f_o(\beta,\theta) + 1$ for the case where $A_{2n+1}$ is used. The required oracle calls in both of the algorithms using the even and odd members then are given by

$$m_e(\beta,\theta) = 2n_e(\beta,\theta) = \begin{cases} \lceil c_e \rceil, & \text{for even } \lceil c_e \rceil, \\ \lceil c_e \rceil + 1, & \text{for odd } \lceil c_e \rceil, \end{cases} \quad (19)$$

and

$$m_o(\beta,\theta) = 2n_o(\beta,\theta) + 1 = \begin{cases} \lceil c_o \rceil, & \text{for odd } \lceil c_o \rceil, \\ \lceil c_o \rceil + 1, & \text{for even } \lceil c_o \rceil, \end{cases} \quad (20)$$

respectively. We show in Figs. 1-3 the particular oracle-call functions for the given $\beta = 10^{-3}, 10^{-1}$, and $10^0$, respectively. For the sake of completeness, in these figures we have also shown the oracle-call function $c = f$ when $G^n$ is used in the search problem, where $f$ is given by [3][4]

$$f = \frac{\frac{\pi}{2} - \sin^{-1}(\sin(\frac{\theta}{2})\sin(\beta))}{2\sin^{-1}(\sin(\frac{\theta}{2})\sin(\beta))}, \quad \text{for } \phi = \theta \quad (21)$$

It is clear that, if given $\beta$, all the functions shown in (19)-(21) have minimal values at $\theta = \pi$, since then their derivatives with respect to $\theta$ will be zero in all the cases. So if minimal oracle calls are demanded in a search problem, one should choose the phase angle $\theta = \theta_{op}$, which is nearest to $\pi$ and satisfies

$$c_e(\beta,\theta_{op}) = m_e(\beta,\pi) \quad \text{for the case using } A_{2n},$$

$$c_o(\beta,\theta_{op}) = m_o(\beta,\pi) \quad \text{for the case using } A_{2n+1},$$

or

$$c(\beta,\theta_{op}) = \lceil f(\beta,\pi) \rceil \quad \text{for the case using } G^n.$$

For example, if given $\beta = 1$, we have $m_e(\beta,\pi) = 4$, $m_o(\beta,\pi) = 1$, and $\lceil f(\beta,\pi) \rceil = 1$, so the phase angles $\theta_{op} = \pi \pm 1.304$, $\pi \pm 1.87$, and $\pi \pm 1.87$ should be chosen in the algorithms using $A_{2n}$, $A_{2n+1}$, and $G^n$, respectively. These choices for the sure-success search have been schematically depicted by the empty circles shown in Fig. 3.

To summarize, we have in this work provided the matching condition $\phi = -\theta$ for the sure-success search algorithm using the even member $A_{2n}$ and the matching condition $\phi = \theta$ for the algorithm in which the even member $A_{2n+1}$ is applied. We have also given the exact formulae for evaluating the required iterations, or oracle calls, in both cases. We have, therefore, formulated the whole family

{$A_n$, $n=1,2,\cdots$} introduced by Hu[5].An example has been given to interpret how to choose the phase angle $\theta_{op}$ if minimal oracle calls are demanded in the sure-success algorithms.

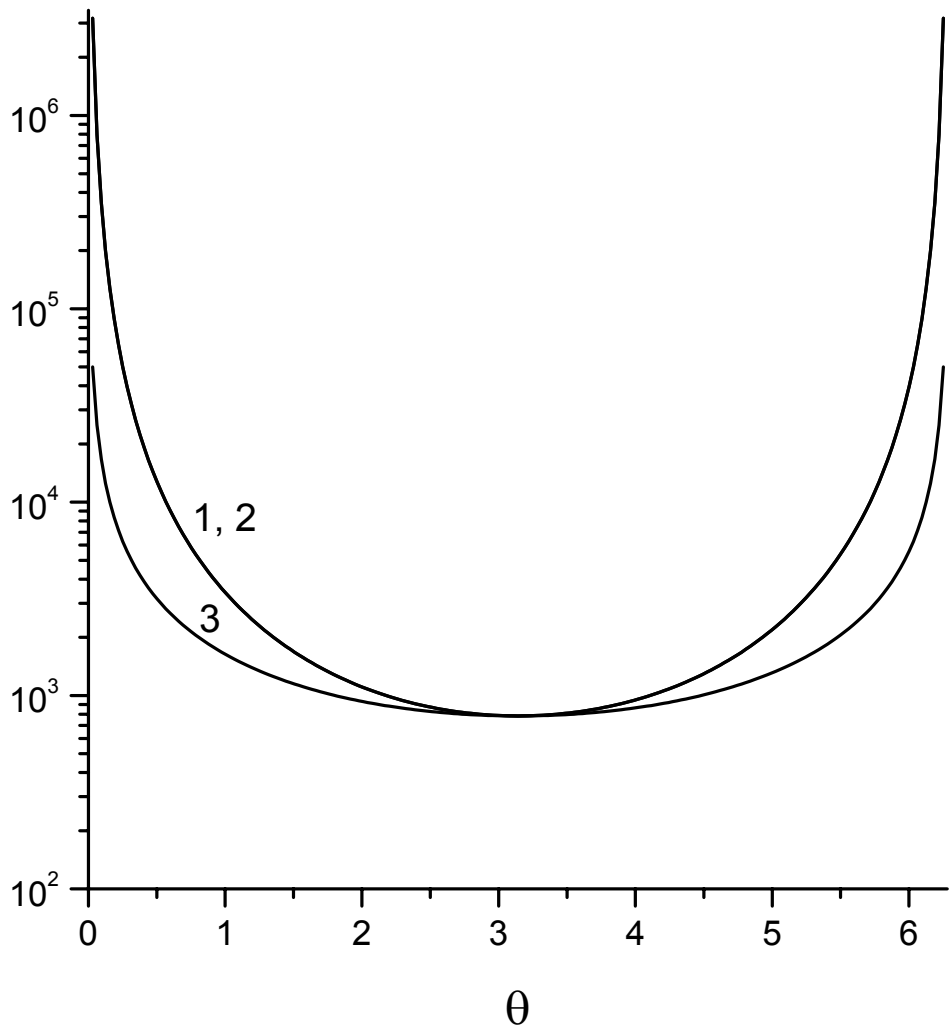

Fig.1. Oracle-call functions for $\beta = 10^{-3}$. Curve 1, 2, and 3 associate with the functions $c_e(\theta)$, $c_o(\theta)$, and $c(\theta)$, respectively. In this case, the curves 1 and 2 are very close.

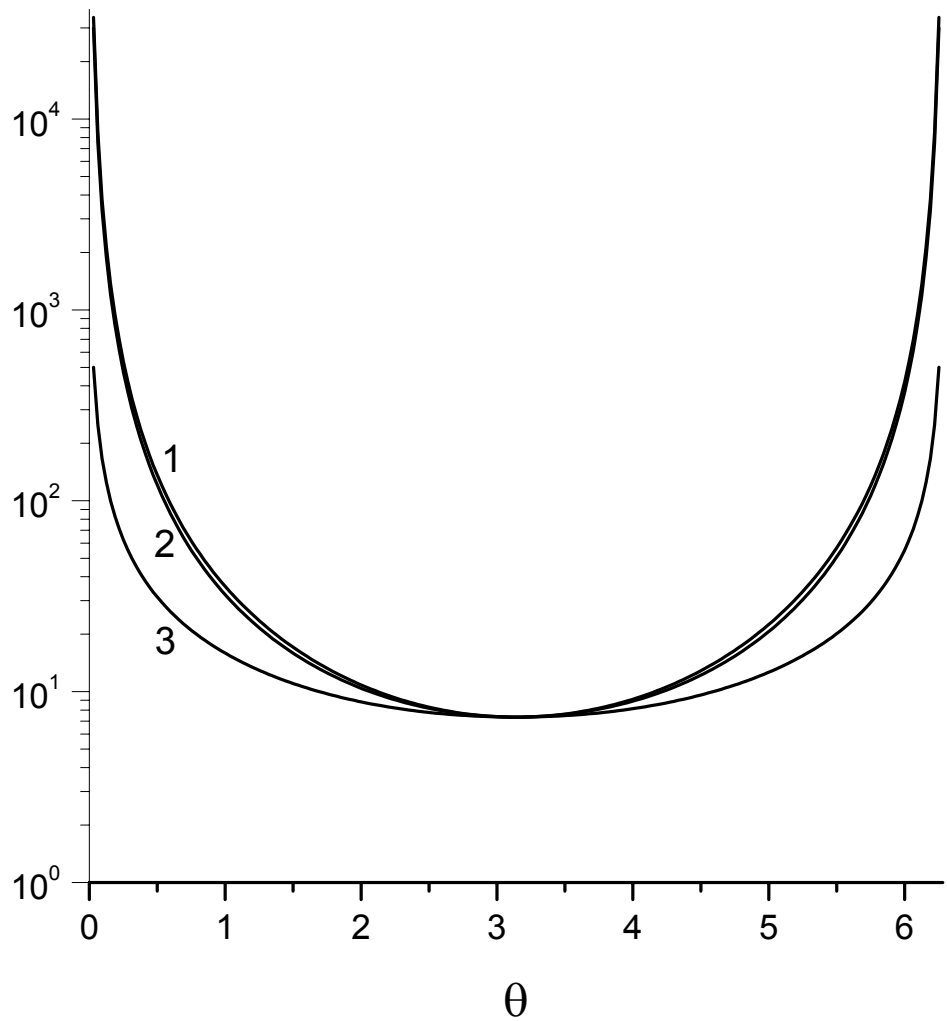

Fig.2. Oracle-call functions for $\beta = 10^{-1}$. Curve 1, 2, and 3 represent the functions $c_e(\theta)$, $c_o(\theta)$, and $c(\theta)$, respectively. In this case, the curves 1 and 2 are also very close.

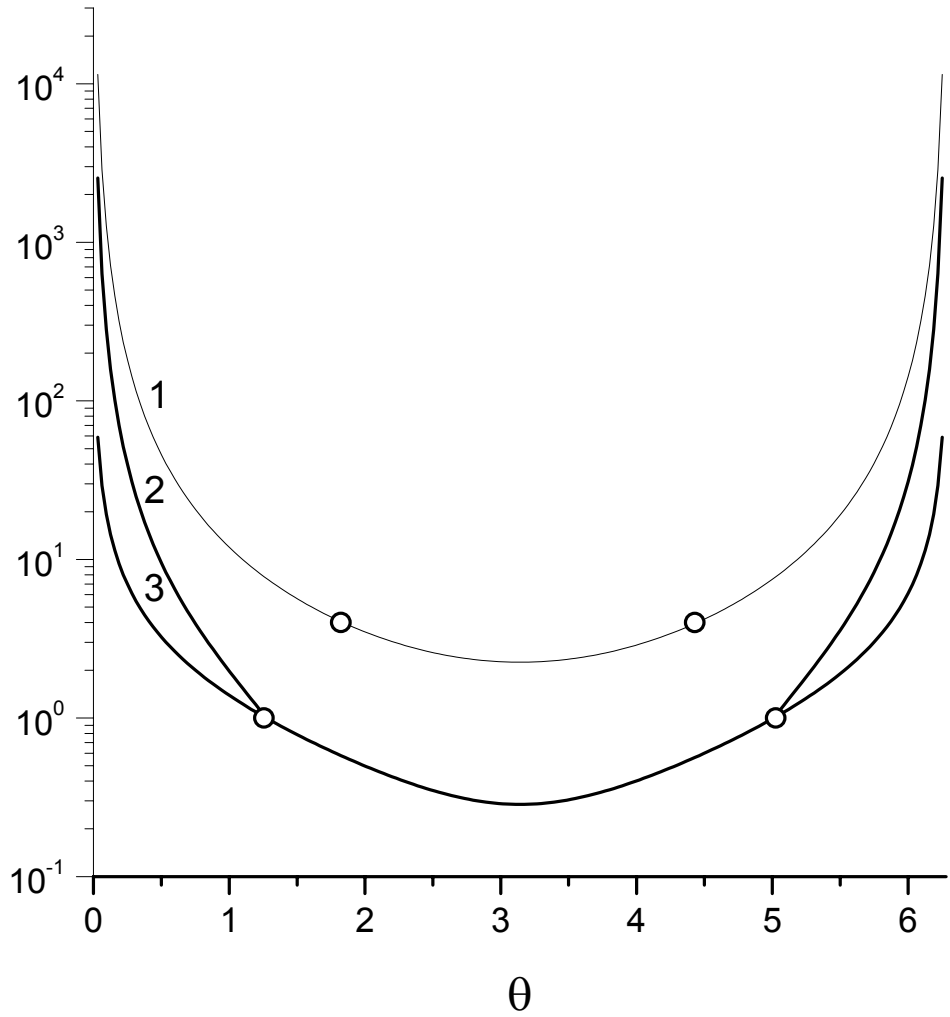

Fig.3. Oracle-call functions for $\beta = 10^0$. Curve 1, 2, and 3 represent the functions $c_e(\theta)$, $c_o(\theta)$, and $c(\theta)$, respectively. The empty circles stand for the cases when minimal oracle calls are demanded.